\documentclass{cimento}
\usepackage{graphicx}
\begin{document}

\title{Effective Field Theory for Top Quark Physics}
\author{Cen Zhang and Scott Willenbrock\from{ins:illinois}}
\instlist{\inst{ins:illinois} Department of Physics, University of Illinois at Urbana-Champaign, 1110 West Green Street, Urbana, IL  61801}


\maketitle

\begin{abstract}
Physics beyond the standard model can affect top-quark physics indirectly.  We describe the effective field theory approach to describing such physics, and contrast it with the vertex-function approach that has been pursued previously.  We argue that the effective field theory approach has many fundamental advantages and is also simpler.
\end{abstract}

There are two broad categories of approaches to physics beyond the standard model (SM). The first is to look for the new physics directly via the production of new particles. The second is to look for new interactions of the known particles of the SM. Here we address the question of the best way to formalize the latter approach \cite{Zhang:2010}.

\begin{figure}[h]
\centering
\includegraphics[scale=1.7]{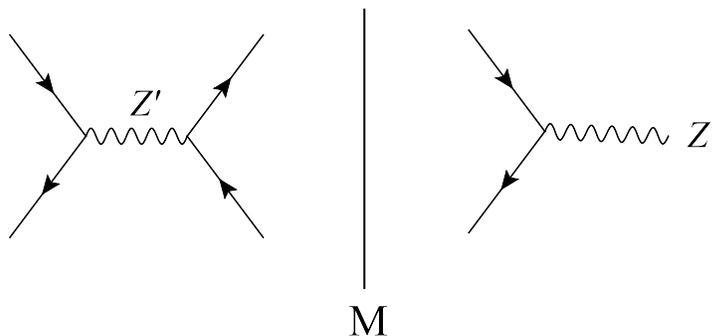}
\caption{At energies greater than the $Z'$ mass, one observes the new particle directly.  At energies below the $Z'$ mass, one observes its effects on SM particles indirectly.}\label{fig:Zprime}
\end{figure}

An example of the two approaches is displayed in fig.~\ref{fig:Zprime}, in the context of a $Z'$ boson. At energies above the mass of the $Z'$, one observes the new particle directly. At energies below the mass of the $Z'$, one observes new interactions of SM fermions mediated by the exchange of a virtual $Z'$ boson. At energies much less than the $Z'$ mass, this appears as an effective four-fermion interaction.

Let us attach a coupling $g$ to the $Z'$ interaction with the SM fermions, and include the $Z'$ propagator, proportional to $(p^2-M^2)^{-1}$. At energies much less than the $Z'$ mass, we can neglect the momentum of the $Z'$ in the propagator. The effective four-fermion interaction is thus of strength $g^2/M^2$, and the theory is described by the Lagrangian
\begin{equation}
	\label{eq:fourfermion}
	\mathcal{L_\mathit{eff}}=\mathcal{L}_{\sy{SM}}+\frac{g^2}{M^2}\bar{\psi}\psi\bar{\psi}\psi\;.
\end{equation}

We can generalize the above discussion with the help of dimensional analysis. Recall that in units where $\hbar=c=1$, the fields of the SM have mass dimensions
\begin{eqnletter}
	\tx{dim\ } A_\mu&=&1\\
	\tx{dim\ } \phi&=&1\\
	\tx{dim\ } \psi&=&3/2
\end{eqnletter}
where $A_\mu$ is a gauge field, $\phi$ is the Higgs field, and $\psi$ is a fermion field.  Every term in $\mathcal{L}_\sy{SM}$ is of dimension four or less, while the new four-fermion interaction is of dimension six. This makes sense, because it has a coefficient with a dimension of two inverse powers of mass.

The generalization of eq.~(\ref{eq:fourfermion}) is now clear:
\begin{equation}
	\mathcal{L_\mathit{eff}}=\mathcal{L}_\sy{SM}+\sum_i\frac{c_i}{\Lambda^2}O_i
\end{equation}
where the sum is over all dimension-six operators (that is, products of SM fields). The coefficient $c_i$ is the coupling of the SM fields to the new physics and $\Lambda$ is the scale at which the new physics resides. This construction, called an effective field theory, is completely general and independent of the details of the new physics, whether it be new particles, extra dimensions, strings, \textit{etc} \cite{Weinberg:1978kz}.

The bad news is that there are over 60 dimension-six operators, not even taking into consideration three generations of fermions \cite{Leung:1984ni,Buchmuller:1985jz,AguilarSaavedra:2009mx}. The good news is that only a few of these operators affect top quark physics. Let's look at an example.

\begin{figure}[h]
\centering
\includegraphics{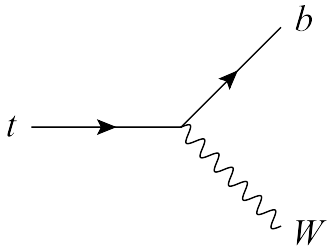}
\caption{Top quark decay.  The $W$ boson goes on to decay to a pair of quarks or leptons.}\label{fig:topdecay}
\end{figure}

The decay of the top quark in the SM is shown in fig.~\ref{fig:topdecay}. The $W$ boson goes on to decay to leptons or light quarks, and we will assume this is described to a good approximation by the SM, since this vertex is well tested.  The $Wtb$ vertex is much less constrained, so we seek a dimension-six operator that affects the decay of the top quark.

In the SM, the branching ratio of the top quark to zero helicity, negative helicity (left handed), and positive helicity (right handed) $W$ bosons is
\begin{eqnletter}
	F_0&=&\frac{m_t^2}{m_t^2+2M_W^2}\approx 0.7\\
	F_L&=&\frac{2M_W^2}{m_t^2+2M_W^2}\approx 0.3\\
	F_R&\approx&0
\end{eqnletter}
where I have neglected the $b$ quark mass, which is a good approximation. Of the more than 60 dimension-six operators, there is only one that affects these branching ratios:
\begin{equation}
	O_{tW}=(\bar{q}\sigma^{\mu\nu}\tau^It)\tilde{\phi}W_{\mu\nu}^I+h.c.
\end{equation}
where $W_{\mu\nu}^I$ is the SU(2) field-strength tensor, $\phi$ is the Higgs doublet, $t$ is the right-chiral top quark, and $q$ is the left-chiral quark doublet containing top and bottom.  The matrix $\sigma^{\mu\nu}=\frac{i}{2}[\gamma^\mu,\gamma^\nu]$ is a tensor constructed from Dirac matrices, and $\tau^I$ are the SU(2) Pauli matrices.

The presence of the Higgs doublet field deserves special attention. We are assuming that the new physics, at scale $\Lambda$, is above the scale of electroweak symmetry breaking, which in the SM is the Higgs boson mass. Thus the Higgs doublet is part of the theory below $\Lambda$, and the dimension-six operators must respect the unbroken $\tx{SU(2)}\times\tx{U(1)}_\sy{Y}$ symmetry of the SM.

When the Higgs field acquires a vacuum-expectation value, the dimension-six operator $O_{tW}$ yields the effective interaction
\begin{equation} \mathcal{L_\mathit{eff}}=\mathcal{L}_\sy{SM}+\frac{C_{tW}}{\Lambda^2}v\left(\bar{b}\sigma^{\mu\nu}(1+\gamma_5)t\right)
	\partial_\mu W_\nu^-+h.c.
\label{eq:tW}
\end{equation}
This unfamiliar interaction is not present in the SM. Its effect is to modify the top-quark branching ratios:
\begin{eqnletter}
F_0&=&\frac{m_t^2}{m_t^2+2m_W^2}-\frac{4\sqrt{2}C_{tW}v^2}{\Lambda^2}\frac{m_tm_W(m_t^2-m_W^2)}{(m_t^2+2m_W^2)^2}\\
F_L&=&\frac{2m_W^2}{m_t^2+2m_W^2}+\frac{4\sqrt{2}C_{tW}v^2}{\Lambda^2}\frac{m_tm_W(m_t^2-m_W^2)}{(m_t^2+2m_W^2)^2}\\
F_R&=&0
\end{eqnletter}
By measuring these branching ratios, we may extract (or place a bound on) the coefficient $C_{tW}/\Lambda^2$.

\begin{figure}[h]
\centering
\includegraphics[scale=1.2]{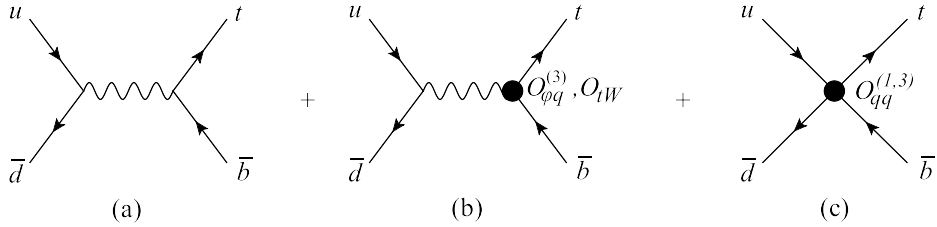}
\caption{Single-top production via the electroweak interaction.  The diagrams as shown correspond to $s$-channel production; $t$-channel production corresponds to the same diagrams turned on their sides.}\label{fig:singletop}
\end{figure}

Let's now turn to another physical process involving the top quark: single top production, shown in fig.~\ref{fig:singletop}. In addition to the operator $O_{tW}$, two other operators affect the single-top cross section \cite{Cao:2007ea}:
\begin{eqnletter}
	O_{\phi q}^{(3)}&=&i(\phi^\dagger \tau^I D_\mu \phi)(\bar{q} \gamma^\mu \tau^I q)+h.c.\\
	O_{qq}^{(1,3)}&=&(\bar{q}^i\gamma_\mu\tau^I q^j)(\bar{q}\gamma^\mu\tau^I q)
\end{eqnletter}
where $D_\mu$ is the gauge-covariant derivative, and the superscripts on the quark doublet fields indicate that they are from the first two generations ($i,j=1,2$). The superscript (3) on the operators indicates that they contain SU(2) triplet currents, and the additional superscript (1) indicates that the currents are color singlets.
When the Higgs field acquires a vacuum-expectation value, the operator $O_{\phi q}^{(3)}$ yields
\begin{equation}
	\mathcal{L_\mathit{eff}}=\mathcal{L}_\sy{SM}+\frac{C_{\phi q}^{(3)}gv^2}{2\sqrt{2}\Lambda^2}
	\left(\bar{b}\gamma^\mu(1-\gamma_5)t\right)W^-_\mu+h.c.
\label{eq:phiq}
\end{equation}
This is identical in form to the $V-A$ coupling of the SM. Thus this operator simply renormalizes the SM charged-current interaction. The operator $O_{qq}^{(1,3)}$ is a four-fermion interaction that couples light quarks to third-generation quarks, and thus contributes to single-top production as shown in fig.~\ref{fig:singletop}(c).

To extract (or bound) the coefficients of the three operators $O_{tW}$, $O_{\phi q}^{(3)}$, and $O_{qq}^{(1,3)}$, the following strategy can be employed.  First, we use top decay to extract $C_{tW}/\Lambda^2$.  Then we use single-top production to extract $C_{\phi q}^{(3)}/\Lambda^2$ and $C_{qq}^{(1,3)}/\Lambda^2$.  The latter coefficient enters with the opposite sign in $s$-channel and $t$-channel single-top production, so it is useful to measure the two processes separately.  In practice, one might fit all three coefficients to all three processes simultaneously.

Let us compare this program with the program that has been followed up until now, which I will refer to as the vertex-function approach.  In this approach, one writes down the most general vertex function, consistent with Lorentz invariance, coupling a virtual $W$ boson to on-shell top and bottom quarks \cite{Kane:1991bg,Boos:1999dd,Chen:2005vr,AguilarSaavedra:2006fy,AguilarSaavedra:2008gt}:
\begin{equation}
\Gamma^\mu_{Wtb} = -\frac{g}{\sqrt 2}V_{tb}\left(\gamma^\mu [f_1^LP_L + f_1^RP_R]-i\frac{\sigma^{\mu\nu}}{M_W}(p_t-p_b)_\nu [f_2^LP_L + f_2^RP_R]\right)
\end{equation}
where $P_{L,R}=(1\mp \gamma_5)/2$.  The four form factors $f_1^L,f_1^R,f_2^L,f_2^R$ are arbitrary functions of the virtuality of the $W$ boson, $Q^2$.  The vertex function is sometimes expressed as a Lagrangian,
\begin{equation}
\mathcal{L}_{Wtb} = \frac{g}{\sqrt 2}V_{tb}\left(W^-_\mu\bar b\gamma^\mu [f_1^LP_L + f_1^RP_R]t-\frac{1}{M_W}\partial_\nu W^-_\mu\bar b\sigma^{\mu\nu}[f_2^LP_L + f_2^RP_R]t\right)
\end{equation}
but this is misleading, because it is not an effective Lagrangian in the modern sense described above.  In addition, the coefficients $f_1^L,f_1^R,f_2^L,f_2^R$ in the Lagrangian cannot be functions of the $W$ boson virtuality, because the Lagrangian is expressed in position space, not momentum space.

In the effective field theory approach, the form factors of the vertex function are derived from the dimension-six operators.  One finds
\begin{eqnletter}
f_1^L = 1+C_{\phi q}^{(3)}\frac{v^2}{\Lambda^2} \\
f_2^R = \sqrt 2 C_{tW}\frac{v^2}{\Lambda^2}\;.
\end{eqnletter}
As mentioned above, the operator $O_{\phi q}^{(3)}$ simply renormalizes the $V-A$ coupling of the SM, as shown in eq.~(\ref{eq:phiq}).  The operator $C_{tW}$ contributes to the right-handed tensor coupling, as shown in eq.~(\ref{eq:tW}).  The other two form factors, $f_1^R$ and $f_2^L$, contribute to physical processes only at order $\mathcal{O}(m_b)$ and $\mathcal{O}(1/\Lambda^4)$, and are hence suppressed.  We neglect them throughout.  Thus, in contrast to the vertex-function approach, the effective field theory approach provides a rationale for neglecting some of the form factors, and also justifies setting the other form factors to constants independent of $Q^2$.

Let us compare the effective field theory approach and the vertex-function approach in more detail.  There are several aspects to the comparison:
\begin{itemize}
\item The effective field theory approach is well motivated and provides guidance as to the most likely places to observe the indirect effects of physics beyond the standard model.  The vertex-function approach does not share either of these features.
\item The effective field theory approach incorporates the known $\tx{SU(3)}_\sy{C}\times\tx{SU(2)}\times\tx{U(1)}_\sy{Y}$ symmetry of the SM.  The vertex-function approach turns its back on gauge symmetry.
\item The effective field theory approach includes contact interactions (such as $O_{qq}^{(1,3)}$) as well as interactions that contribute to vertices.  The vertex-function approach only concerns itself with the latter.
\item The effective field theory approach is valid regardless of whether the top and bottom quarks are real or virtual.  The vertex-function approach assumes that the top and bottom quarks are on shell.  If they are off shell, one must add additional form factors.
\item In the effective field theory approach, one can calculate radiative corrections associated with SM vertices as well as with those arising from the dimension-six operators.  There is no unambiguous way to calculate radiative corrections in the vertex-function approach.
\end{itemize}
This last point may be the most important of all.  We aspire to make contact between direct top-quark measurements and precision electroweak measurements, where the top quark enters in loops.  Because the effective field theory is renormalizable in the modern sense, it is possible to carry this program out.  All divergences that arise from dimension-six operators can be absorbed in the coefficients of other dimension-six operators.  So although the theory is not renormalizable in the old-fashioned sense, it allows for the unambiguous calculation of radiative corrections to $\mathcal{O}(1/\Lambda^2)$.

\begin{figure}[h]
\centering
\includegraphics[scale=1.25]{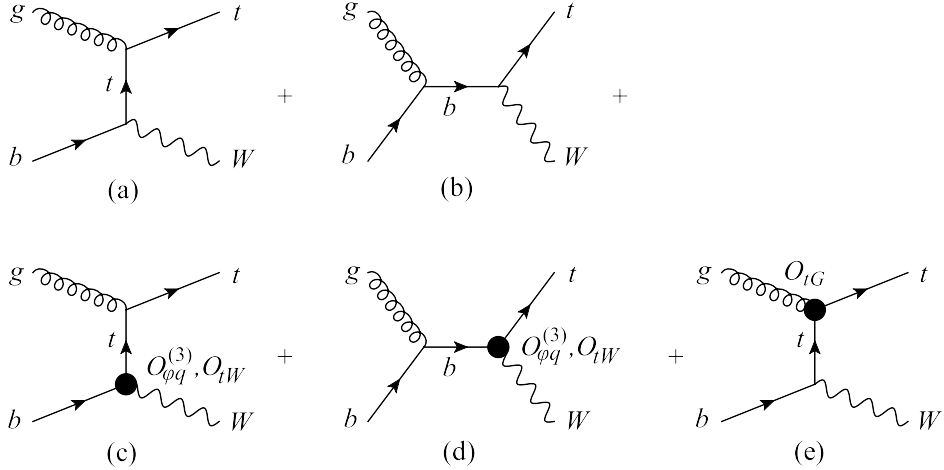}
\caption{Associated production of a single top quark and a $W$ boson.}\label{fig:Wt}
\end{figure}

Let us pick up where we left off and ask what other dimension-six operators can be probed by top-quark physics.  We show in fig.~\ref{fig:Wt} another single-top process, where the top quark is produced in association with a $W$ boson.  This involves two of the same operators, $O_{tW}$ and $O_{\phi q}^{(3)}$, that we encountered in $s$- and $t$-channel single top production, along with a new operator:
\begin{equation}
	O_{tG}=(\bar{q}\sigma^{\mu\nu}\lambda^At)\tilde{\phi}G_{\mu\nu}^A+h.c.
\end{equation}
where $\lambda^A$ is an SU(3) matrix and $G_{\mu\nu}^A$ is the gluon field-strength tensor.  This operator is the strong-interaction analogue of $O_{tW}$.  The four-quark operator $O_{qq}^{(1,3)}$ does not contribute to this process.  By measuring $Wt$ associated production, together with the other weak interaction processes discussed above, we can extract (or bound) the coefficient $C_{tG}/\Lambda^2$.

\begin{figure}[h]
\centering
\includegraphics[scale=0.86]{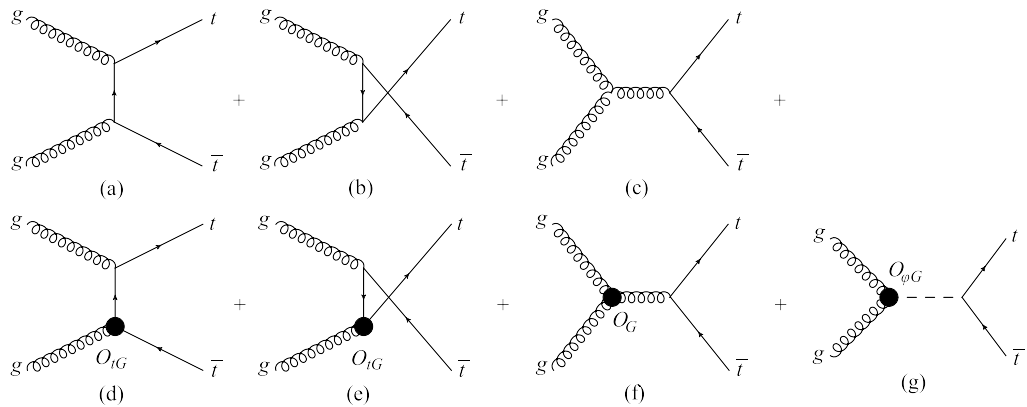}
\caption{Top-quark pair production via a gluon-sluon collision.}\label{fig:ggtt}
\end{figure}

Finally, let's consider top-quark pair production.  At the LHC, this is dominated by the subprocess $gg\to t\bar t$, as shown in fig.~\ref{fig:ggtt}.  This is influenced by the operator $O_{tG}$ that we just discussed, but also by the operators
\begin{eqnletter}
	O_G &=&f_{ABC}G_\mu^{A\nu} G_\nu^{B\rho} G_\rho^{C\mu}\\
    O_{\phi G}&=&\frac{1}{2}\phi^\dagger\phi G_{\mu\nu}^AG^{A\mu\nu}
\end{eqnletter}
The operator $O_G$ affects the triple-gluon vertex.  Although it does not involve the top quark directly, bounds on this operator from other processes are rather weak, so the best measurement (or bound) may come from top-quark pair production \cite{Cho:1994yu}.  The operator $O_{\phi G}$ couples the Higgs field directly to a pair of gluons, and will be measured (or bounded) by Higgs production \cite{Manohar:2006gz}.  Other operators are probed by $q\bar q\to t\bar t$ \cite{Cho:1994yu}; I leave that to another presenter \cite{qqtt}.

\newpage
{\bf Summary} -- The time is ripe to abandon the vertex-function approach and to adopt the more modern and systematic approach of effective field theory for top-quark physics.  Although effective field theories have been around for a long time and are widely used in particle physics, nuclear physics, condensed matter physics, and elsewhere, they have been slow to be adopted in the context of anomalous couplings.  While I was already convinced of the timeliness of the effective field theory approach prior to this workshop, my conviction was further solidified on my arrival day when I noticed a popular Belgian beer, $\mathcal{L\!\mathit{effe}}$, that beckons us to embrace effective Lagrangians.


\acknowledgments
This work was supported in part by the
U.~S.~Department of Energy under contract No.~DE-FG02-91ER40677.


\begin{thebibliography}{0}

\bibitem{Zhang:2010} C.~Zhang and S.~Willenbrock, work in progress.

\bibitem{Weinberg:1978kz}
  S.~Weinberg,
  Physica A {\bf 96}, 327 (1979).

\bibitem{Leung:1984ni}
  C.~N.~Leung, S.~T.~Love and S.~Rao,
  Z.\ Phys.\  C {\bf 31}, 433 (1986).

\bibitem{Buchmuller:1985jz}
  W.~Buchmuller and D.~Wyler,
  Nucl.\ Phys.\  B {\bf 268}, 621 (1986).

\bibitem{AguilarSaavedra:2009mx}
  J.~A.~Aguilar-Saavedra,
  Nucl.\ Phys.\  B {\bf 812}, 181 (2009)
  [arXiv:0811.3842 [hep-ph]].

\bibitem{Cao:2007ea}
  Q.~H.~Cao, J.~Wudka and C.~P.~Yuan,
  Phys.\ Lett.\  B {\bf 658}, 50 (2007)
  [arXiv:0704.2809 [hep-ph]].

\bibitem{Kane:1991bg}
  G.~L.~Kane, G.~A.~Ladinsky and C.~P.~Yuan,
  Phys.\ Rev.\  D {\bf 45}, 124 (1992).

\bibitem{Boos:1999dd}
  E.~Boos, L.~Dudko and T.~Ohl,
  Eur.\ Phys.\ J.\  C {\bf 11}, 473 (1999)
  [arXiv:hep-ph/9903215].

\bibitem{Chen:2005vr}
  C.~R.~Chen, F.~Larios and C.~P.~Yuan,
  Phys.\ Lett.\  B {\bf 631}, 126 (2005)
  [AIP Conf.\ Proc.\  {\bf 792}, 591 (2005)]
  [arXiv:hep-ph/0503040].

\bibitem{AguilarSaavedra:2006fy}
  J.~A.~Aguilar-Saavedra, J.~Carvalho, N.~F.~Castro, F.~Veloso and A.~Onofre,
  Eur.\ Phys.\ J.\  C {\bf 50}, 519 (2007)
  [arXiv:hep-ph/0605190].

\bibitem{AguilarSaavedra:2008gt}
  J.~A.~Aguilar-Saavedra,
  Nucl.\ Phys.\  B {\bf 804}, 160 (2008)
  [arXiv:0803.3810 [hep-ph]].

\bibitem{Cho:1994yu}
  P.~L.~Cho and E.~H.~Simmons,
  Phys.\ Rev.\  D {\bf 51}, 2360 (1995)
  [arXiv:hep-ph/9408206].

\bibitem{Manohar:2006gz}
  A.~V.~Manohar and M.~B.~Wise,
  Phys.\ Lett.\  B {\bf 636}, 107 (2006)
  [arXiv:hep-ph/0601212].

\bibitem{qqtt}
C. DeGrande, J.-M.~Gerard, C.~Grojean, F.~Maltoni, and G.~Servant, these proceedings

\end{thebibliography}
\end{document}